\documentclass[letter,12pt]{article}
\pdfoutput=1 

\usepackage{jheppub} 

\usepackage[utf8]{inputenc}
\usepackage[english]{babel}

\usepackage[T1]{fontenc} 

\usepackage{comment}

\def\bea{\begin{eqnarray}}
\def\eea{\end{eqnarray}}

\title{ Chiral transport in curved spacetime via holography }

\author[a]{Alexander Avdoshkin,}
\author[b,c]{Rustem Sharipov}

\affiliation[a]{Department of Physics, University of California, Berkeley, CA, USA 94720\\}
\affiliation[b]{Russian Quantum Center, Skolkovo, Moscow 143025, Russia\\}
\affiliation[c]{Moscow Institute of Physics and Technology,\\Institutskii per, 9, 141700, Dolgoprudny, Russia\\}

\emailAdd{alexander\_avdoshkin@berkeley.edu}
\emailAdd{sharipov.ro@phystech.edu}

\abstract{We consider a holographic model of strongly interacting plasma with a gravitational anomaly.
In this model, we compute parity-odd responses of the system at finite temperature and chemical potential to external electromagnetic and gravitational fields. Working within the linearized fluid/gravity duality, we performed the calculation up to the third order in gradient expansion. Besides reproducing the chiral magnetic (CME) and vortical (CVE) effects we also obtain gradient corrections to the CME and CVE due to the gravitational anomaly. Additionally, we find energy-momentum and current responses to the gravitational field similarly determined by the gravitational anomaly. The energy-momentum response is the first purely gravitational transport effect that has been related to quantum anomalies in holographic theories.}

\begin{document}
\maketitle
\flushbottom

\section{Introduction}
\indent \indent Gauge and gravitational quantum axial anomalies are a peculiar feature of quantum field theories with fundamental chiral fermions. First discovered in relation to neutral pion decay\cite{Adler, Bell:1969ts}, they are manifested in a non-conservation of the classically conserved axial current at the quantum level in the presence of external gauge and gravitational fields\cite{Adler, GAUME}:

\begin{equation} \label{anomalies}
\partial_{\mu} j_{L/R}^{\mu} = \kappa \epsilon^{\mu\nu\lambda\rho} F_{\mu \nu} F_{\lambda\rho} + \lambda \epsilon^{\mu\nu\lambda\rho} R^{\alpha}_{\beta \mu \nu} R^{\beta}_{\alpha \lambda\rho},
\end{equation}
where $F^{\mu\nu}$ is the electromagnetic tensor, $R_{\mu\nu \alpha \beta}$ is the Riemann tensor, $ \epsilon^{\mu\nu\lambda\rho}$ is the Levi-Civita symbol, and $\kappa$, and $\lambda$ are theory-dependent parameters. For a single left Dirac fermion of unit charge,  $\kappa = \frac{e^2}{32\pi ^2}$ and $\lambda = \frac{1}{768 \pi ^2}$ . Axial anomalies are robust in the sense that Eq. (\ref{anomalies}) is exactly given by the one-loop contribution and does not depend on the energy scale in the theory \cite{AdlerBardeen}.

Recently, anomalies have been shown to lead to a special class of transport phenomena – chiral effects. Assuming that a system is in an external magnetic field and has 
local velocity $u_{\rho}$, the parity-odd parts of the expectation values of the axial $j^{\mu}_{A}$ and vector $j^{\mu}_{V}$ currents are given by \cite{Kharzeev_2016}
\begin{eqnarray} \label{currents}
(j^{\mu}_{V})_{odd} =  \sigma_{B}^{(V)} B^{\mu} + \sigma_{V}^{(V)} \omega^{\mu},~ (j^{\mu}_{A})_{odd} =  \sigma_{B}^{(A)} B^{\mu} + \sigma_{V}^{(A)} \omega^{\mu},
\end{eqnarray}
where $\omega^{\mu} = \epsilon^{\mu \nu \rho \sigma}u_\nu \nabla_\rho u_\sigma $ is the vorticity of the fluid flow and $B^{\mu}=\frac{1}{2}\epsilon^{\mu \nu \rho \sigma} u_\nu F_{\rho \sigma}$ is the magnetic field in the local rest frame. The vector and axial currents along $B^\mu$
are referred to as the chiral magnetic (CME) and chiral separation effects (CSE), correspondingly, and the currents along $\omega^{\mu}$ are chiral vortial effects (CVE). For a single Dirac fermion of unit charge, $\sigma_{B}^{(A/V)} = \frac{\mu_{V/A} }{2 \pi^2}$, $\sigma^{(A/V)}_{V} = \frac{ \mu_A \mu_V }{2 \pi^2}$, $\xi^{(V)}_{A} = \frac{\mu_A^2  + \mu_V^2}{4\pi^2} + \frac{T^2}{12}$, where $\mu_{A/V}$
is the chemical potential conjugated to the axial/vector charge\cite{Kharzeev_2016}. Although, the CME seems to allow for a non-zero equilibrium electric current one should be warranted from this interpretation as it has been shown to be a purely dynamical response \cite{Vazifeh}.

Instances of chiral effects have been discussed in various areas of physics. The original derivation was done by Vilenkin \cite{vilenkin, plasmas} with applications for cosmology/astrophysics. Later, the significance of chiral effects was also realized for the quark-gluon plasma regime of QCD \cite{qcd} where they are believed to be responsible for the charge dependence of the hadron elliptic flow \cite{heavy}. 
In solid state physics, chiral effects are relevant for the physics of Weyl and Dirac semimetals \cite{Burkov_2015, Hosur_2013}, where chiral fermions are realized as low-energy excitations. Experimentally, the observed negative magneto resistance in $ZrTe_5$ has been linked to the presence of the CME \cite{Kharzeev_2018, negative} and the thermoelectrical transport in NbP was found to be consistent with the gravitational anomaly \cite{Gooth_2017}. Chiral effects were also considered in cold atoms \cite{atoms}. 

There are several ways in which chiral effects modify the dynamics of a system. In \cite{instab} it was shown that a system with chiral imbalance is unstable toward spontaneous generation of helical magnetic field due to chiral effects. For a more detailed discussion of instabilities see \cite{tuchin2019timeevolution, PhysRevD.92.074018, PhysRevD.94.025009, PhysRevD.92.125031, Avdoshkin_2016, Miransky_2015}. Additionally, in strong external magnetic field chiral media exhibit anomalous class of novel excitations \cite{wave, Shovkovy_2019,Mottola:2019nui}.

A direct connection of the CME and the CSE to the axial anomaly has been discussed in literature, for review see e.g. \cite{Kharzeev_2016}. Connection in  the hydrodynamic approximation was first discussed in \cite{surowka} and in generalization with two charges \cite{Sadofyev:2010pr}.  On the other hand, the thermal part of CVE is not determined by the chiral gauge anomaly \cite{oz}. In \cite{Landsteiner_2011}, the authors argued that it is the gravitational chiral anomaly that is responsible for the thermal part of the CVE and this relation has been established via the mixed anomaly in the dual five-dimensional theory. A discussion of the direct connection of this effect to the gravitational anomalies can be found in \cite{Stone_2018, Jensen_2013, Golkar_2015, Hou_2012, Glorioso_2019, PhysRevD.98.096011,Avkhadiev:2017fxj,Prokhorov:2020okl} and references therein.

In this work, we explore further connections between anomalies and transport responses within the framework of holography. In particular, extending the analysis of \cite{Meg_as_2013, Bu:2015ika}, we will show that the chiral gravitational effect ($\chi_{T}$ in \ref{gravtens}), reported in \cite{Ma_es_2013, Jensen_2013, Sadofyev_2018}, is in fact determined by the gravitational anomaly in holographic theories. We also find that the CME and CVE ($\xi_B$ and $\xi_V$ in Eq. (\ref{vector_response})) receive gradient corrections that are determined by the gravitational anomaly (previously analysed numerically in \cite{Landsteiner:2013aba}). And, finally, we found a novel odd current response to gravity ($\chi_{V}$ in \ref{gravvec}) that is similar to the energy-momentum response $\chi_T$. The contributions independent of the gravitational anomaly agree with the results reported earlier, see e.g. \cite{Bu_2019}. 

Similarly to \cite{Landsteiner_2011}, we will consider a strongly interacting holographic plasma in the presence of a mixed gauge-gravitational anomaly. According to the AdS/CFT correspondence, in the strongly coupled large-N regime this model is dual to a 5D Einstein-Maxwell theory in AdS \cite{Maldacena_1999}. Finite temperature and chemical potential at the boundary are accounted for by a black hole in the bulk. The axial anomaly in the R-charged current is present in $\mathcal{N}=4$ SYM and its effect on the transport coefficients has been considered in, e.g., \cite{Yee_2010, Erdmenger_2009}. The gravitational anomaly is not normally present in $\mathcal{N}=4$ SYM and we introduce it via a mixed anomaly in the bulk \cite{Meg_as_2013, Landsteiner_2011}. This model was previously considered in \cite{Meg_as_2013}, where the effects of the gravitational anomaly on flat-space hydrodynamic responses were computed for up to the second order in gradients. We consider the linearized fluid/gravity correspondence in curved background and calculate responses up to the third order in gradients.

The work is structured as follows. In Section \ref{section:responses} we  review the general structure of the linear hydrodynamic, identify all the relevant responses and present our results. In Section \ref{section:model} we describe the holographic model and outline the procedure for the calculation of transport coefficients. In Section \ref{section:solution} we solve the holographic equations perturbatively and relate the solutions to the responses of the boundary theory while more technical details are presented in the appendix. Finally, in Section \ref{section:discussion} we conclude with a discussion of the results. 
\section{Linear Responses in Hydrodynamics and Results} \label{section:responses}
 \indent  \indent   Here we introduce the transport effects discussed above within the hydrodynamics framework \cite{Kovtun_2012}. Let us begin by considering a relativistic many-body system in an equilibrium state with temperature $T$ and chemical potential $\mu$ moving with constant velocity $u^{\mu}$. We assume that the system has only one (anomalously) conserved current of  right- or left-handed fermions.  In this  case, the expectation values of the energy-momentum tensor and current are given by 
\bea \label{eq_emt}
\left<T^{\mu\nu}\right> = (\epsilon + p)u^{\mu} u^{\nu} + p \eta^{\mu \nu},~ \left<J^{\mu}\right> = n u^{\mu}, 
\eea
where $P$ and $\epsilon$ are functions of $T$ and $\mu$ corresponding to the local pressure and energy density of the system and $\eta^{\mu\nu}$ is the flat Minkowski metric. 

Eq. (\ref{eq_emt}) can be generalized to states that are at equilibrium only locally and, thus, are described by slowly varying $T(x^{\mu})$, $\mu(x^{\mu})$, and $u^{\mu}(x^\nu)$ \cite{Kovtun_2012}. Additionally, we allow for curved backgrounds described by metric tensor $g^{\mu\nu}(x^{\mu})$ which we also assume to be slowly varying. Now the expectation values in Eq. (\ref{eq_emt}) take form

\begin{subequations}\label{constituitve}
\begin{equation}
\left<T^{\mu\nu}\right> = \mathcal{E}u^{\mu} u^{\nu} + \mathcal{P} \Delta^{\mu \nu}+\left( q^\mu u^\nu+q^\nu u^\mu \right)+ \tau^{\mu \nu },
\end{equation}
\begin{equation}
\left<J^{\mu}\right> = \mathcal{N} u^{\mu}+\nu^\mu,
\end{equation}
\end{subequations}

where $\Delta^{\mu \nu}=g^{\mu \nu}+u^\mu u^\nu$ is the projector on the directions transverse to the fluid velocity vector, and $\mathcal{E},\mathcal{P},\mathcal{N}$, $q^\mu$, $\tau^{\mu\nu}$, and $\nu^\mu$ are functions of $T$, $\mu$, $u^\mu$. In order to make the decompositions unique we also require that $q^\mu u_\mu=\nu^\mu u_\mu=0 $ and the tensor $\tau^{\mu\nu}$ be transverse (with respect to the velocity vector), symmetric and traceless. The definitions of local $u^{\mu}(x)$, $T(x)$ and $\mu(x)$ are made precise by working in the Landau frame, where we require $q^\mu=0$, $\mathcal{E}=\epsilon$ and $\mathcal{N}=n$ \cite{Kovtun_2012}.

Because we only confine ourselves to conformal system in this work the stress-energy tensor has to satisfy an additional constraint \cite{Erdmenger_2009}
\begin{equation}
\left< T_{\mu}^\mu \right>=0,
\end{equation}
which in equilibrium is equivalent to equation of state $\epsilon=3 p$. Whereas, the presence of an anomalous current obeying Eq. (\ref{anomalies}) necessitates an inclusion of parity-odd terms in the constitutive relations Eq. (\ref{constituitve}), we will separate those contributions by decomposing $\tau^{\mu\nu}= \tau^{\mu\nu}_{\text{odd}} + \tau^{\mu\nu}_{\text{even}},~~\nu^{\mu} = \nu_{\text{odd}}^{\mu}+ \nu_{\text{even}}^{\mu}$, where $\tau^{\mu\nu}_{\text{even}}$,  $\nu_{\text{even}}^{\mu}$ are invariant under spatial parity and $\tau^{\mu\nu}_{\text{odd}}$,  $\nu_{\text{odd}}^{\mu}$ change sign.  Additionally, we only focus on the part of $\tau_{odd}^{\mu\nu}$ and $\nu_{odd}^{\mu}$ that is linear in the amplitude of deviations from the equilibrium. The most general expression for $\tau_{odd}^{\mu\nu}$ and $\nu_{odd}^{\mu}$ reads as
\begin{subequations}\label{odd_constitutive}
\begin{equation}\label{tensor_response}
 \tau^{\mu\nu}_{\text{odd}}=\chi_T~\Delta^{e  < \mu }\epsilon^{\nu> \beta c d} u_{\beta} \nabla_c R_{d e} + \mathcal{O}(\text{deviation}^2),
 \end{equation}
 \begin{equation}\label{vector_response}
  \nu_{odd}^\mu=\xi_V~ \omega^\mu+\xi_{B}~B^\mu+\chi_V~\epsilon^{\mu \nu \rho \sigma} u_\nu \nabla_\rho  R_{\sigma \alpha} u^\alpha + \mathcal{O}(\text{deviation}^2),
\end{equation}
\end{subequations}
 where $R_{\mu \nu}$ is the Ricci tensor correspondning to $g_{\mu \nu}$ and $\xi_{B/V}$, $\chi_{V/T}$ are functions of derivative operators $u^\mu \nabla_\mu$ and $\nabla^\mu \nabla_\mu$. Also, for arbitrary rank-2 tensor $A_{\mu \nu}$ we have introduced
$$
A_{< \mu \nu >}=\Delta^{\lambda}_{\mu} \Delta^{\sigma}_{\nu}A_{\lambda \sigma}-\frac{1}{3} \Delta_{\mu \nu} \Delta^{\lambda \sigma} A_{\lambda \sigma}.
$$


Earlier in Eq. (\ref{currents}), the coefficients $\xi_{B}$ and $\xi_V$ have been identified as the Chiral Magnetic/Separation Effects and Chiral Vortical Effects correspondingly. $\xi_T$ is the fluid response to gravity that have been observed in \cite{Ma_es_2013, Jensen_2013}  and $\chi_V$ is novel gravitational response in the conserved current.

So far the discussion has been compeletely general and Eqs. (\ref{odd_constitutive}) apply to any model with appropriate symmetries. Now we present our results for a holographic model with gauge and gravitational anomalies described in detail in Section \ref{section:model}. This theory has three free parameters: $G_5$, $\bar{\lambda}$ and $\bar{\kappa}$ that correspond to the number of degrees of freedom of the theory and strengths of the gauge and gravitation anomalies, respectively. Hereinafter, we will consider expressions linearized in $\mu /T$. For the equations of state of this system we obtain
\begin{equation}
\epsilon=3 p=3 \frac{(\pi T)^4}{16 \pi G_5}~~~~~\frac{n}{\epsilon+p}=\frac{\mu}{2(\pi T)^2}.
\end{equation}

The transport coefficients in Eqs. (\ref{odd_constitutive}) have been computed up to the third order in gradients in the static limit ($u^\mu \nabla_\mu = 0$). In Fourier space  transport coefficients become functions of  $k^2$ and read
\begin{subequations} \label{result}
\begin{gather}
16 \pi G_5~\xi_B=16 \bar{\lambda} \mu-8  \bar{\kappa} \mu -\nonumber \\ 
-\left[\left( \pi^2+2\right)\bar\lambda-\frac{\pi^2}{3} \bar\kappa-1344 \bar\kappa \bar\lambda^2+7872 \bar\lambda^3\right] \mu \left(\frac{k}{\pi T} \right)^2+\mathcal{O} (k^3)
\end{gather}\begin{equation}
16 \pi G_5 ~ \xi_V= -32  \bar\lambda \pi^2 T^2 + \frac{2}{3}  \bar\lambda k^2 \left[ \pi^2-6~\log(2)+6336  \lambda^2 \right]+ \mathcal{O} (k^3)
\end{equation}and for gravitational transport coefficients
\begin{equation}\label{gravtens}
16 \pi G_5 ~\chi_T=-16 \bar{\lambda} \mu +\mathcal{O} (k)
\end{equation}
\begin{equation}\label{gravvec}
16 \pi G_5 ~\chi_V=-8\bar{\lambda} \left(\log(2)-1 \right)+\mathcal{O} (k)
\end{equation}

\end{subequations}

	The $k^2$ correction to $\xi_B$ was computed in \cite{Bu_2019} and coincides with our result in the absence of the gravitational anomaly. Whereas, all contributions involving $\bar{\lambda}$ were not previously known analytically.
	
	By virtue of Eq. (\ref{lambda}) the transport coefficients in Eq. (\ref{result}) are funcions of only the anomaly coefficient defined in Eq. (\ref{anomalies}) : $\chi_T=16 \lambda \mu +\mathcal{O} (k) $ and $
\chi_V=8 \lambda \left(\log(2)-1 \right)+\mathcal{O} (k) $ which suggests that they might have a universal relation to the anomalies of the corresponding quantum field theory. This is corroborated by the fact that our results for $\chi_T$ agree with those computed for other models \cite{Jensen_2013, Ma_es_2013}. \\

\section{The Holographic Model} \label{section:model}

\indent  \indent  We work with the four-dimensional theory dual to the Einstein-Maxwell theory with gauge and mixed anomalies in AdS$_5$.  The equilibrium state of the boundary theory at finite temperature and finite chemical potential corresponds to the bulk solution with a charged black hole. To study the response to an external gravitational field we need to find the bulk solution where the boundary limit of the gravitational field corresponds to perturbed metric.

The model we will use was introduced in \cite{Landsteiner_2011}. The signature of the five-dimensional metric is chosen to be $(-,+,+,+,+)$. The Levi-Civita tensor is defined by  $\epsilon_{MNPQR} = +\sqrt{-g} \epsilon(MNPQR)$, where $\epsilon(MNPQR$)  is a totally anisymmetric tensor normalized by $\epsilon(01234)=+1$. The main part of the bulk action reads
\begin{eqnarray}\label{action}
S_{EM} = \frac{1}{16 \pi G_5}\int_{\mathcal{M}} d^5 x \sqrt{-g} \left[ R + 2 \Lambda - \frac{1}{4}F_{MN}F^{MN} \right.+\nonumber\\ 
+\left.\epsilon^{MNPQR}A_{M}\left(\frac{\bar{\kappa}}{3}F_{NP}F_{QR}+\bar{\lambda} R^{A} {}_{BNP} R^{B} {}_{AQR}\right) \right],
\end{eqnarray}
where $R$ is the Ricci scalar of the bulk metric $g_{MN}$, $\Lambda$ is the cosmological constant that determines the size of the AdS radius: $\Lambda=\frac{6}{L_{AdS}^2}$. We  choose units such that $L_{AdS}=1$ and $\Lambda=6$.
This model reduces to the $\mathcal{N} = 4$ SYM when we put $\bar{\kappa} = \frac{1}{4 \sqrt{3}}$ and $\bar{\lambda} = 0$. 

Since our manifold has a boundary, the Gibbons–Hawking–York boundary term $S_{GHY}$ has to be added to make the variational problem well-defined. In the case of action Eq. (\ref{action}), it takes the form 
\begin{equation}\label{actionbndr}
S_{GHY}=\frac{1}{8 \pi G_5} \int_{\partial M}d^4 x \sqrt{-\gamma} K-\frac{1}{2 \pi G_5} \int_{\partial M}d^4 x \sqrt{-\gamma} \lambda n_M \epsilon^{MNPQR} A_N K_{PL} D_Q K^L_R,
\end{equation}
where $n_A$ is a unit vector normal to the holographic boundary $\partial \mathcal{M}$, $D_A$ is covariant derivative induced on the boundary, and $\gamma^{\alpha \beta}$ is the metric induced on the fixed $r$ hypersurface. We have also used the extrinsic curvature  $K_{\mu \nu}=\gamma^\alpha_\mu D_{\alpha} n_\nu$ and its trace $K = \gamma^{\mu \nu} K_{\mu \nu}$. The second boundary term is required to reproduce the gravitational anomaly at general boundary hypersurface.

The renormalization of action Eq. (\ref{action}) is achived by the introduction of counterterms:
\begin{equation}\label{actionct}
S_{ct}=\frac{1}{16 \pi G_5} \int_{\partial M} d^4 x \sqrt{-\gamma} \left[6+\frac{1}{2}\hat{R} -\log \left(\frac{1}{r^2} \right) \left(\frac{1}{8}\hat{R}^{\mu \nu} \hat{R}_{\mu \nu}-\hat{R}^2 -\frac{1}{8} \hat{F}^2\right)\right]
\end{equation}
where $\hat{R}_{\mu \nu}$ and $\hat{F}_{\mu \nu}$  are the four dimensional Ricci tensor and field strength induced on the boundary. The total action of the theory is  $S=S_{EM}+S_{GHY}+S_{ct}$.


\normalsize

Action Eq. (\ref{action}) results in the following bulk equations of motion
\begin{eqnarray}\label{eom}
G_{MN} - \Lambda g_{MN} = \frac{1}{2} F_{ML}F_{N}^{L} - \frac{1}{8}F^2 g_{MN} + 2 \bar{\lambda} \epsilon_{LPQR(M}\nabla_B (F^{PL} R^B_{N)} {}^{QR} )\\
\nabla_N F^{NM} = - \epsilon^{MNPQR} (\bar{\kappa} F_{NP}F_{QR} + \bar{\lambda} R^{A}_{BNP} R^{B}_{AQR}),\nonumber
\end{eqnarray}
where $G_{MN}$ is the Einstein tensor. The boundary metric of the dual conformal theory is related to $\gamma$ as $\lim\limits_{r \rightarrow \infty}\left(\gamma_{\mu \nu} / r^2\right) = g_{\mu \nu}^{CFT}$. The consistent boundary stress-energy tensor is defined as 
\begin{equation}\label{T}
T^{\mu \nu}_{(con)}=-\lim_{r \rightarrow \infty} \left( r^2 \frac{2}{\sqrt{-\gamma}} \frac{ \delta S }{\delta \gamma_{\mu \nu}}\right)
\end{equation}

In order to make it gauge invariant one has to add the Chern-Simons current, for a more detailed discussions see \cite{Meg_as_2013}. In particular, for our model from (\ref{action}),(\ref{actionbndr}) and (\ref{actionct}) we obtain covariant stress-energy tensor 
\begin{eqnarray}\label{TEM}
T_{\mu \nu}=\frac{1}{16 \pi G_5}\lim_{r \rightarrow \infty} \left[ 2 r^2\left( K_{\mu \nu}-K \gamma_{\mu \nu}-3 \gamma_{\mu \nu}- \frac{1}{2} \hat{G}_{\mu \nu}\right) +T^{ct}_{\mu \nu} \log \left( \frac{1}{r^2}\right)+\right. \nonumber\\
+2 \bar\lambda \epsilon_{(\mu \alpha \beta \rho}F^{\alpha \beta} R_{\nu)}^{\rho} \Bigr]
\end{eqnarray}
where $\hat{G}_{\mu \nu}$ is the Einstein tensor induced on the boundary. The term with $T_{\mu \nu}^{ct}$, obtained by variation of counterterm part of the action, removes logarithmic divergences in the presence of an external metric. As is explained in \cite{Bu:2015ika}, logarithms only contribute to fourth and higher orders in derivatives expansion that lie beyond our consideration.

The consistent current is defined as the variation of the action with respect to the source

\begin{equation}
J^\mu_{(con)}= \lim_{r \rightarrow \infty} \frac{\delta S}{\delta A_\mu}
\end{equation}
It is related to the gauge invariant covariant current by addition of a Chern-Simons term. The explicit expression for the covariant boundary current is

\begin{equation}\label{CURR}
J^{\mu}=\frac{1}{16 \pi G_5} \lim_{r \rightarrow \infty } \sqrt{-\gamma} \left[F^{\mu r} + \frac{1}{2} D_\alpha F^{\alpha\mu} \log \left( \frac{1}{r^2} \right)\right].
\end{equation}

Using Maxwell's equations, one can get the following non-conservation of the (covariant) boundary current

\begin{equation}
D_\mu J^\mu=-\frac{1}{16\pi G_5}\epsilon^{\mu\nu\lambda\rho}\left(  \bar\kappa  F_{\mu \nu} F_{\lambda\rho} + \bar\lambda  R^{\alpha}_{\beta \mu \nu} R^{\beta}_{\alpha \lambda\rho}\right)
\end{equation}

This corresponds to Eq. (\ref{anomalies}) with
\begin{eqnarray}\label{lambda}
\kappa = - \frac{\bar\kappa}{16 \pi G_5} ~~ ,~ \lambda = - \frac{\bar{\lambda}}{16 \pi G_5}.
\end{eqnarray}

The equilibrium configuration of the boundary theory corresponds to the Reissner-Nordstrom black brane solution of equations of motion (\ref{eom}), which in the Eddington-Finkelstein coordinates reads 
\begin{subequations}\label{BlackHole}

\begin{equation}
ds^2 = g^{(0)}_{M N}d x^{M} d x^{n} = 2 dt dr-r^2 f(r)dt^2+r^2 (dx^i)^2
\end{equation}
\begin{equation}
A = -\frac{\sqrt{3}Q}{r^2}dt,
\end{equation}
\end{subequations}
where $f(r)=1-\frac{M}{r^4}+\frac{Q^2}{r^6}$,  $M$ is the mass and $Q$ is the charge of the black hole. Using the $AdS/CFT$ dictionary \cite{Erdmenger_2009,Cvetic:1999ne}, we can relate the characteristics of the black hole with the equilibrium parameters of the boundary theory:
\begin{gather}
M=\frac{(\pi T)^4}{2^4} \left(1+\sqrt{1+\frac{2 \mu^2}{3 \pi^2 T^2}} \right)^3\left(3\sqrt{1+\frac{2 \mu^2}{3 \pi^2 T^2}} -1\right), \\
Q=\frac{\mu \pi^2 T^2 }{4\sqrt{3}}\left(1+\sqrt{1+\frac{2 \mu^2}{3 \pi^2 T^2}} \right)^2.
\end{gather}

The solution Eq. (\ref{BlackHole}) allows a generalization to the boosted version of a black hole

\begin{subequations}\label{blackhole}
\begin{equation}
ds^2=-r^2 f(r) u_\mu u_\nu dx^\mu dx^\nu+r^2 \Delta_{\mu \nu}^{(0)}dx^\mu dx^\nu -2u_\mu dx^\mu dr,
\end{equation}
\begin{equation}
A_{\mu}=\frac{\sqrt{3}Q u_{\mu}}{r^2},
\end{equation}
\end{subequations} 
where $\Delta_{\mu \nu}^{(0)}=\eta_{\mu \nu}+u_\mu u_{\nu}$, $u_{\mu}$ is constant and normalized by $u^2=-1$.
Now we consider the perturbations of the background metric and gauge field, closely following \cite{Bu:2015ika}.
In order to reproduce the fluid dynamics of the boundary theory, we let the velocity be a function of  the coordinates. The curved metric on the boundary is introduced by defining the projector transverse to velocity field  with respect to background metric $\Delta_{\mu \nu}(x)=g^{CFT}_{\mu \nu}(x)+u_{\mu }(x)u_{\nu}(x)$. Additionally, we modify the guage field in the bulk to yield $a^{CFT}(x)$ on the boundary. With all the above modifications, the perturbed black brane metric reads
\begin{subequations}\label{blackhole2}
\begin{equation}
ds^2=-r^2 f(r) u_\mu(x) u_\nu(x) dx^\mu dx^\nu+r^2 \Delta_{\mu \nu}(x)dx^\mu dx^\nu -2u_\mu(x) dx^\mu dr,
\end{equation}
\begin{equation}
A_{\mu}=\frac{\sqrt{3}Q u_{\mu}(x)}{r^2}+a^{CFT}_\mu (x),
\end{equation}
\end{subequations} 
where the velocity field has to be normalized with respect to the curved metric
\begin{equation}
g^{CFT}_{\mu\nu}(x) u^\mu(x) u^\nu(x)=-1.
\end{equation}

Metric (\ref{blackhole2})  has the desired boundary behaviour with $\lim\limits_{r\rightarrow \infty}  \left( g_{\mu \nu}(x,r)/r^2 \right) =g^{CFT}(x)$.

 Following the approach outlined in the previous section, we will consider small perturbations of the dual theory parameters and write
\begin{subequations}\label{perturb}
\begin{equation}
g_{\mu \nu }^{CFT}(x)=\eta_{\mu \nu} +h^{CFT}_{\mu \nu},
\end{equation}\begin{equation}
u_{\mu}(x)=\left(-1+\frac{1}{2}h_{00}^{CFT}(x),v_i(x) \right),
\end{equation}
\end{subequations}
with $h^{CFT}_{\mu \nu}$ and $v_{\mu}$ taken to be small.
    The linearized version of (\ref{blackhole2}) (that we will call the seed metric) is
\begin{subequations}\label{seed}
\begin{equation}
ds^2_{seed}=2 dt dr-r^2 f(r) dt^2 +r^2 dx_i^2- \left[ 2 v_i(x)dr dx^i+\frac{2}{r^2}v_i(x)dt dx^i\right.\nonumber
\end{equation}
\begin{equation}
 \left.+h_{00}^{CFT}(x) dr dt +\frac{1}{r^2}h_{00}^{CFT}(x)dt^2-r^2 h_{\mu \nu}^{CFT}(x)dx^\mu dx^\nu \right]
\end{equation}
\begin{equation}
A_{seed}= -\frac{\sqrt{3}Q}{r^2}dt+ \left[\frac{\sqrt{3}Q}{2 r^2}h_{00}^{CFT}(x)dt+ \frac{\sqrt{3}Q}{ r^2}v_{i}(x) dx^i+a_\mu^{CFT}(x) dx^\mu\right]
\end{equation}
\end{subequations}

The seed metric and gauge field  (\ref{seed}) reproduce the correct perturbations of boundary theory, but do not satisfy the Einstein-Maxwell equations Eqs. (\ref{eom}). 
In order to satisfy (\ref{eom}) we write the total the bulk configuration as
\begin{subequations}
\begin{equation}
ds^2 = (ds^2)_{seed} + (ds^2)_{corr},
\end{equation}
\begin{equation}
A = A_{seed} +A_{corr},
\end{equation}
\end{subequations}
where $s_{corr}$ and $A_{corr}$ do not alter the boundary behaviour but ensure that Eqs. (\ref{eom}) are satisfied.
We choose the "background field" gauge for the correction metric:
\begin{equation}
g_{rr}=0 ~,~g_{r \mu} \sim u_\mu ~~, \left( g^{(0)}\right)^{-1}_{MN} g_{corr}^{MN} =0,
\end{equation}

and the axial gauge for the gauge field $A_r=0$. With this choice of the gauges, the most general form of the correction metric and  gauge field is
\begin{subequations}

\begin{equation}\label{metrdec}
ds^2_{corr} = -3 h dr dt+\frac{k}{r^2}dt^2+r^2 h dx_i^2+2 r^2 j_i dt dx^i+r^2 \pi_{ij}dx^i dx^j
\end{equation}
\begin{equation}
A_{corr}= c ~dt+ a_i ~dx^i
\end{equation}\end{subequations}
where $k$, $h$ etc. are functions of radial coordinates and boundary perturbations. The tensor $\pi_{i j}$ is  symmetric and traceless. Since the seed metric should change the boundary perturbations the functions in the decomposition Eq. (\ref{metrdec})  must satisfy:
\begin{equation}\label{bndcnd}
\pi_{ij}=o(r^0)~,~j_i=o(r^0)~,~h=o(r^0)~,~   k=o(r^4)~,~c=o(r^0)~,~a_i=o(r^0).
\end{equation}

The remaining boundary conditions for the functions in Eq. (\ref{metrdec}) are provided by the Landau frame conditions and the requirement of regularity at the black hole horizon. 

 Substituting (\ref{CURR}) for the expectation value of the parity-odd component of the boundary current, we obtain 
\begin{equation}\label{curresp}
16 \pi G_5~ J_i^{odd}=2 a^{(\bar{2})}_i,
\end{equation}where ($\bar2$) denotes the $1/r^2$ term in large $r$ expansion. While exact form of stress-energy tensor of the correction metric  (\ref{metrdec}) is presented in Appendix \ref{section:appB}.

We find the stress-energy tensor and current density of the fluid at rest by substituting the unperturbed solution (\ref{BlackHole}) in Eqs. (\ref{TEM}) and (\ref{CURR}), respectively.  It is the equilibrium stress-energy tensor and current (\ref{eq_emt}) with
\begin{equation}
\epsilon =3 p =\frac{3M}{16 \pi G_5}, ~~n =\frac{\sqrt{3}Q }{8 \pi G_5}.
\end{equation}

\section{Perturbative solution of Einstein equations} \label{section:solution}

\indent  \indent  The equations satisfied by $g_{corr}$ and $A_{corr}$ are the linearized versions of Eqs. (\ref{eom}). It is a set of fifteen equations which can be divided into ten dynamical equations and four constraint equations (one dynamical equation is not independent), for more details see e.g. \cite{Bhattacharyya:2008jc}.  The solution is uniquely determined by the dynamical equations and the Landau frame conditions while the constraint equations are the Navier–Stokes equations of the fluid dynamics that ensure the conservation of the boundary energy-momentum tensor.

We will be working in Fourier space and for that purpose we define
\begin{gather}
g_{M N}(\vec{x}, r)=\int e^{i k_i x^i} g_{M N}(\vec{k},r) d^3 k~,~~A_{\mu}(x,r)=\int e^{i k_i x^i} A_{\mu}(\vec{k},r) d^3k,~\nonumber\\
v_{\mu}(\vec{x})=\int e^{ik_i x^i} v_{\mu}(\vec{k}) d^3k.
\end{gather}

We decompose the correction metric and gauge potential into irreducible representations of $SO(2)$ rotations around $\vec{k}$ \cite{Yee_2010}. The irreducible representations are indexed by helicity and the possible values are $0,\pm 1, \pm 2$ and will be listed below. Without loss of generality we choose $\vec{k}=(0,0,k_3)$. As the equations for different helicities do not mix we study each mode separately. In the remainder of the work, will be working with dimensionless quantities and the physical units can be restored by multiplying by the proper power of $\pi T$.

\subsection{Helicity 0 modes}

\indent  \indent This sector is comprised of 6 modes:
	\begin{equation}
   k~ ,~~h~,~~\pi_{33}=-\left( \pi_{11}+\pi_{22} \right),~j_3~~c ,~~ a_{3}.
    \end{equation}
    
The corresponding equations of motion are independent of the anomaly coefficients and, consequently, there are no parity odd contributions to the transport coefficients from these modes.

\subsection{Helicity $\pm 1$ modes}

\indent  \indent The helicity $\pm1$ modes are:
\begin{equation}
j_{\pm1}=\left( j_{1}\pm i ~j_{2} \right)~, ~~ \pi_{\pm 1}= \left( \pi_{13} \pm i \pi_{23} \right),~~a_{\pm 1}=(a_1 \pm i a_2).
\end{equation}

Similarly, we introduce
\begin{equation}
j^{CFT}_{\pm 1}=h^{CFT}_{01} \pm i h_{02}^{CFT}~,~~\pi^{CFT}_{\pm 1}=h^{CFT}_{13} \pm i h^{CFT}_{23}~,~~v_{\pm 1}= v_1\pm i v_2
\end{equation}

The dynamical equations of this sector are
\begin{subequations}
\begin{itemize}

\item $ \mathcal{M}_1\pm i \mathcal{M}_2:$
\end{itemize}
\begin{equation}
-\frac{1}{r}\left(r^3 f(r) a'_{\pm1} \right)'+\left( \frac{k_3^2}{r^2} \mp \frac{16 \bar\kappa Q k_3}{r^4} \right) a_{\pm 1}+\frac{2 \sqrt{3} Q}{r} j_{\pm 1}'\pm \frac{24 \bar\lambda  k_3 \left(5 Q^2-2 r^2\right) j_{\pm 1}'}{r^5}  +\nonumber 
\end{equation}
\begin{equation}
+ \frac{ k_3 v_{\pm 1} \left(\sqrt{3} k_3 Q  \right) }{r^4} \mp \ k_3 v_{\pm 1}  \frac{48}{r^{12}} \Bigl[ 15 \bar\lambda  Q^4+Q^2 \left(\bar\kappa  r^6-16 \bar\lambda  r^2\right)+4 \bar\lambda  r^4 \Bigr]+\nonumber 
\end{equation}
\begin{equation}\label{M1}
+\left( \frac{k_3^2}{r^2} \mp \frac{16 \bar\kappa Q k_3}{r^4} \right) a_{\pm 1}^{CFT}=0
\end{equation}

\begin{itemize}
\item $ \mathcal{E}_{t1}\pm i \mathcal{E}_{t2}:$
\end{itemize}

\begin{equation}
- \frac{f(r)}{2r}\left( r^5 j'_{\pm 1}\right)'+ \frac{k_3^2}{2} j_{\pm 1}-\frac{\sqrt{3}Q}{r}f(r) a'_{\pm 1}\mp \frac{4 \sqrt{3} \bar\lambda k_3^3 Q }{r^4}j_{\pm 1} \mp\nonumber 
\end{equation}
\begin{equation}
\mp\frac{4 \bar\lambda  k_3}{r^6}  f(r) \Bigl[ 3 \left(2 r^2-5 Q^2 \right)r a_{\pm 1}'+12 a_{\pm 1} \left(5 Q^2-r^2\right)+\sqrt{3} Q 
   r^5 \left(r j_{\pm 1}''+j_{\pm 1}'\right)\Bigr]\mp \nonumber
   \end{equation}
   \begin{equation}
\mp \frac{4 \sqrt{3} k_3 \bar\lambda Q}{ r^{10}} \Bigl[  k_3^2 (-Q^2+r^2)+2r^2 ( 63Q^2-20r^2) f(r)\Bigr] v_{\pm 1}+\frac{Q^2-r^2}{2 r^6} k_3^2 v_{\pm 1}+\nonumber
\end{equation}
\begin{equation}\label{Et1}
\frac{k_3^2}{2}j_{\pm 1}^{CFT}\mp \frac{4 \sqrt{3} \bar\lambda k_3^3 Q }{r^4}j_{\pm 1}^{CFT}\mp\frac{48 \bar\lambda  k_3}{r^6}  f(r)  \left(5 Q^2-r^2\right)a_{\pm 1}^{CFT}=0
\end{equation}

\begin{itemize}
\item $ \mathcal{E}_{13}\pm i \mathcal{E}_{23}:$
\end{itemize}
\begin{equation}
-\frac{1}{2r}\left( r^5 f(r) \pi'_{\pm 1}\right)'+\frac{i k_3 r^2}{2} j'_{\pm 1}+\frac{3 i k_3 r }{2} j_{\pm 1} +\left(\frac{3 i k_3 r}{2}\mp\frac{4 \sqrt{3} \bar{\lambda}Q ik_3^2}{r^3} \right)j_{\pm 1}^{CFT} \pm \nonumber
\end{equation}
\begin{equation}\label{E13}
\pm \frac{4 \sqrt{3} \bar\lambda  k_3 Q }{r^3}\Bigl[ i k_3  \left(r j_{\pm}'-j_{\pm 1}\right)-r^2\left(r  f(r)  \pi_{\pm 1}' \right)'
   \Bigr]-\frac{3}{2}i k_3 r v_{\pm 1}=0,
\end{equation}
and there is a single constraint equation:

\begin{itemize}
\item $r^2 f(r) \left(\mathcal{E}_{r1} \pm i  \mathcal{E}_{r2}\right)+\left(\mathcal{E}_{t1}\pm i \mathcal{E}_{t2} \right):$
\end{itemize}
\begin{equation}\label{Navie}
 k_3  j_{\pm 1}+i f(r) r^2
   \pi'_{\pm 1}+k_3j_{\pm1}^{CFT}-k_3 v_{\pm 1}=0.
\end{equation}

\end{subequations}

It is straightforward to check that Eq. (\ref{E13}) follows from Eqs. (\ref{Et1}) and (\ref{Navie}). Now we solve these equations peturbatively in $Q$ and $k_3$ up to the linear order in $Q$ and third order in $k_3$. Since we are only working to the linear order in $Q$ we can set $f(r)=1-\frac{1}{r^4} $ and, consequently, $r_H=1$ in (\ref{E11}). We represent the functions as series 
\begin{subequations}
\begin{equation}
j_{\pm 1}(r)= \sum_{n=0}^{\infty}j^{(n,0)}_{\pm 1}(r) k_{3}^n+Q \sum_{n=0}^{\infty}j^{(n,1)}_{\pm 1}(r) k_{3}^n~~,
\end{equation}
\begin{equation}
 ~~a_{\pm 1}(r)= \sum_{n=0}^{\infty}a^{(n,0)}_{\pm 1}(r) k_{3}^n+Q \sum_{n=0}^{\infty}a^{(n,1)}_{\pm 1}(r) k_{3}^n.
\end{equation}
\begin{equation}
\pi_{\pm 1}(r)= \sum_{n=0}^{\infty}\pi^{(n,0)}_{\pm 1} +Q \sum_{n=0}^{\infty}\pi^{(n,1)}_{\pm 1}
 \end{equation}
\end{subequations}

We substitute the expansions for $j$ and $a$ into (\ref{M1}), (\ref{Et1}) and solve them order by order. After that $\pi_{\pm 1}$ can be found from Eq. (\ref{Navie}).

Below we list the solutions for the first few orders:

\begin{itemize}
\item Zeroth  order 
\end{itemize}
\begin{equation}
j^{(0,0)}_{\pm 1}(r)=0~~~~~a^{(0,0)}_{\pm 1}(r)=0~~~~
\end{equation}

\begin{itemize}
\item Zeroth order in $Q$ and first order in $k$ 
\end{itemize}
\begin{equation}
j_{\pm 1}^{(1,0)}=\pm 8~ a^{CFT}_{\pm 1} \bar\lambda \frac{1}{r^6}~~~~~~~~a_{\pm 1}^{(1,0)}=\pm  8~ v _{\pm 1} \bar\lambda \left[ \frac{1}{r^4}+2~ \log \left(\frac{1+r^2}{r^2} \right) \right]
\end{equation}

\begin{itemize}
\item First order in $Q$ and first order in $k$
\end{itemize}
\begin{subequations}\begin{equation}
j_{\pm 1}^{(1,1)}=\pm v_{\pm 1} \frac{4 \sqrt{3} ~ \bar\lambda}{r^8} \left[ 3-r^4+2r^6+2r^4(r^4-1)\log\left( \frac{r^2}{1+r^2}\right)\right]
\end{equation}
\begin{equation}
a_{\pm 1}^{(1,1)}=\mp a^{CFT}_{\pm 1}\frac{4 \sqrt{3} ~}{r^4} \left[\bar\lambda +r^4 (\bar\kappa -2 \bar\lambda ) \ln \left(\frac{r^2}{r^2+1} \right) \right]
\end{equation}\end{subequations}

\begin{itemize}
\item Zeroth order in $Q$ and second order in $k$ 
\end{itemize}
\begin{subequations}
\begin{equation}
j_{\pm 1}^{(2,0)}=v_{\pm 1}\frac{1 }{240 r^{10}} \Bigl[ -30 r^8-1024 \bar\lambda ^2 \left(5 r^4 \left(6 r^4-3 r^2+2\right)-9\right) \Bigr]+\nonumber
\end{equation}
\begin{equation}
+v_{\pm 1}\frac{ 5}{80
   r^{6}}  \left[-2 r^2 \coth
   ^{-1}\left(r^2\right)+2048 \bar\lambda ^2 \left(r^6+1\right)\log \left( \frac{r^2+1}{r^2}\right)+r^6 \log\left( \frac{r^2+1}{r^2-1}\right)\right]+\nonumber
\end{equation}
\begin{equation}
+j_{\pm 1}^{CFT}\frac{1}{16 r^4} \left[-2 r^2+\left(r^4-1\right) \log \left(\frac{r^2-1}{r^2+1} \right)\right]
\end{equation}\begin{equation}
a_{\pm 1}^{(2,0)}=-a_{\pm 1}^{CFT} \frac{1}{16 } \left[4 \text{Li}_2\left(r^{-2}\right)-\text{Li}_2\left(r^{-4}\right)+4 \log(r)
   \log \left(\frac{r^2+1}{r^2-1} \right)\right]- \nonumber
\end{equation}
\begin{equation}
-a_{\pm 1}^{CFT}48\bar\lambda ^2 \frac{3 r^4+1}{r^6}
\end{equation}
\end{subequations}
\begin{itemize}
\item First order in $Q$ and second order in $k$  
\end{itemize}

In this and higher orders the expressions for the solutions become too cumbersome and we provide only their asymptotic behavior at $r \to \infty$ 
\begin{subequations}
\begin{equation}
j_{\pm 1}^{(2,1)}\xrightarrow[r\rightarrow \infty]{}~ 0
\end{equation}
\begin{equation}
a_{\pm 1}^{(2,1)}\xrightarrow[r\rightarrow \infty]{} ~\frac{1}{8 r^2} j_{\pm 1}^{CFT}\sqrt{3}
\end{equation}\end{subequations}

\begin{itemize}
\item Zeroth order in  $Q$ and third order in  $k$ 
\end{itemize}
\begin{subequations}
\begin{equation}
j_{\pm 1}^{(3,0)} \xrightarrow[r\rightarrow \infty]{}~0
\end{equation}
\begin{equation}
a_{\pm 1}^{(3,0)}\xrightarrow[r\rightarrow \infty]{}~\pm \frac{2 \bar\lambda }{ r^2}\left(\log(2)-1 \right) j_{\pm 1}^{CFT} \mp \frac{1}{3 r^2}\bar\lambda  v_{\pm 1} \left[\pi ^2-6 \log (2) +6336 \bar\lambda ^2\right]
\end{equation}
\end{subequations}
\begin{itemize}
\item First order in  $Q$ and third order in $k$
\end{itemize}
\begin{subequations}
\begin{equation}
j_{\pm 1}^{(3,1)}\xrightarrow[r\rightarrow \infty]{}~0
\end{equation}
\begin{equation}
a_{\pm 1}^{(3,1)}\xrightarrow[r\rightarrow \infty]{} ~\pm \frac{\sqrt{3}}{r^2}Q k_3^3 a^{CFT}_{\pm1} \left[\frac{\left( \pi^2+2\right)}{2}\bar\lambda- \bar\kappa\frac{\pi^2}{6}+672 \bar\kappa \bar\lambda^2-3936 \bar\lambda^3\right] 
\end{equation}\end{subequations}

 We did not present the solutions for $\pi_{\pm 1}$ as they do not contribute to either the boundary energy momentum tensor or current.
\subsection{Helicity $\pm 2$ modes}

\indent  \indent  This sector contains two independent modes with opposite helicities
\begin{eqnarray}
\pi_{\pm 2} =  ( \pi_{11}- \pi_{22} \pm 2i  \pi_{12})
\end{eqnarray}
each governed by a single equation $(\mathcal{E}_{11}-\mathcal{E}_{22} \pm 2 i \mathcal{E}_{12} )=0:$
\begin{gather}\label{E11}
-\frac{1}{2 r}\left( r^5 f(r) \pi'_{\pm 2}\right)'+\frac{k^2_3}{2} \pi_{\pm 2}\pm \frac{8 \sqrt{3} \bar\lambda  k_3 Q }{r^4}\left[ r^3 \left(\pi _{\pm 2}' r f(r)\right)'-k_3^2  \pi_{\pm 2}\right]+\nonumber\\
+\left[\frac{k^2_3}{2} \pm \frac{8 \sqrt{3} \bar\lambda  k_3^3 Q }{r^4} \right] \pi_{\pm 2}^{CFT}=0.
\end{gather}

Analogously to the previous subsection, we solve the equations order by order in $k_3$ and $Q$. We expand $\pi_{\pm 2}$ in series 

\begin{equation}
\pi_{\pm 2}(r)= \sum_{n=0}^{\infty}\pi^{(n)}_{\pm 2} k_{3}^n~~~,
\end{equation}

Here we list the solutions at each order:
\begin{itemize}
\item Zero and first orders in $k$
\end{itemize}

\begin{equation}
\pi^{(0)}_{\pm 2}(r)=\mathcal{O}(Q^2),~~~~\pi_{\pm 2}^{(1)}=\mathcal{O}(Q^2)
\end{equation}

\begin{itemize}
    \item  Second order in $k$
\end{itemize}
\begin{equation}
\pi^{(2)}_{\pm 2}(r)=\frac{1}{4} \pi_{\pm 2}^{CFT} \log \left(\frac{r^2}{r^2+1}\right)+\mathcal{O}(Q^2)
\end{equation}
\begin{itemize}
\item Third order in $k_3$
\end{itemize}
\begin{eqnarray}
\pi^{(3)}_{\pm 2}=\pm \pi_{\pm 2}^{CFT}\frac{4 \sqrt{3}  \bar\lambda  Q }{r^4}\left[4 r^2+4 r^4 \log \left(\frac{r^2}{r^2+1}\right)-1\right]+\mathcal{O}(Q^2).
\end{eqnarray}

 \subsection{Holographic constitutive relations } \label{subsection:constitutive}
\indent  \indent      In this subsection we use Eqs. (\ref{SET}) and (\ref{curresp}) to evaluate the boundary  expectation values that correspond to the solution obtained above.
    
    The total odd response coming the helicity $\pm 1$ sector (Eq. (\ref{curresp})) is
    \begin{equation}
   \pm  8 \pi G_5 J_{\pm 1}^{(odd)}=\nonumber
    \end{equation}\begin{equation}
  =  4 \sqrt{3} Q (\bar\kappa-2\bar\lambda) k_3 a^{CFT}_{\pm 1} + \sqrt{3}Q k_3^3 a^{CFT}_{\pm1} \left[\frac{\left( \pi^2+2\right)}{2}\bar\lambda- \bar\kappa\frac{\pi^2}{6}-672 \bar\kappa \bar\lambda^2+3936 \bar\lambda^3\right] +\nonumber
    \end{equation}\begin{equation}
    + 16 k_3v_{\pm 1} \bar\lambda - \frac{1}{3} k_3^3\bar\lambda  v_{\pm 1} \left[\pi ^2-6 \log (2) +6336 \bar\lambda ^2\right] -\nonumber
    \end{equation}
    \begin{equation}\label{J1}
 -2 \bar\lambda  k_3^3 \pi_{\pm 1}^{CFT} \left( \ln 2-1\right)+\mathcal{O}(k_3^4), 
    \end{equation}
    where we have introduced $J_{\pm 1}=J_1\pm i J_2$.
    
The Landau frame conditions require $T_{\pm 1} = 0$.

    Having defined $T_{\pm 2}=T_{11}-T_{22}\pm 2 i T_{02}$, for the odd repsonse in the helicity $\pm 2$ sector we obtain 
 \begin{eqnarray}\label{J2}
4 \pi G_5 T_{\pm 2}^{(odd)} =   \pm  4 \sqrt{3} k_3^3 Q \bar\lambda  \pi_{ \pm 2}^{CFT}+\mathcal{O}(k_3^4).
\end{eqnarray}

Matching these expressions with the linearized versions of the covariant expressions in (\ref{odd_constitutive}), ones arrives at the transport coefficients in Eqs. (\ref{result}). 
\section{Discussion} \label{section:discussion}
\indent  \indent In this work, we have studied the implications a gravitational anomaly has on the transport in a holographic model. This approach had the advantage of having the anomaly coefficients as free parameters of the model and, thus, allowed us to establish a more direct relation between the anomalies and transport coefficients. Having extended the linearized treatment of the holographic model with a gravitational anomaly in the fluid/gravity regime described in \cite{Bu:2015ika} to include external gauge fields, we have determined all parity-odd responses up to the third order in gradients. We have reproduced the CME and CVE as well as their previously known gradient correction due to the gauge anomaly \cite{Bu_2019}. Along with that, we found gradient corrections to the CME and CVE coming from the gravitational anomaly with the correction to the CME being particularly interesting as it shows that the gravitational anomaly can be probed through purely electromagnetic responses. Finally, we observed an energy-momentum response to the gravitational field that was discussed earlier but not in a holographic setup. Its dependence on the strength of the gravitational anomaly is consistent with its value reported for free fermions \cite{Ma_es_2013} and suggests its universal relation to the anomaly. Our results show that the impact of the gravitational anomaly on transport goes beyond the $ T ^ 2 $ term in the CVE. The approach developed in this work can be used to study the effects of the gravitational anomaly to higher orders in gradients and to analyse the potential instabilities due to the gravitational chiral effects.

\section{Acknowledgements}
\indent  \indent The authors are grateful to A. Sadofyev for suggesting the problem. We also thank  V.I. Zakharov and  G. Prokhorov for comments on a draft of this paper. The reported 
study was funded by RFBR according to the research project 18-02-40056. A.A. acknowledges support from NSF grant DMR-1918065 and a Kavli ENSI fellowship at UC Berkeley.


\bibliographystyle{JHEP}
\bibliography{grav_anom}

\providecommand{\href}[2]{#2}\begingroup\raggedright\begin{thebibliography}{10}

\bibitem{Adler}
S.~L. Adler, \emph{Axial-vector vertex in spinor electrodynamics},
  \href{https://doi.org/10.1103/PhysRev.177.2426}{\emph{Phys. Rev.} {\bfseries
  177} (1969) 2426}.

\bibitem{Bell:1969ts}
J.~Bell and R.~Jackiw, \emph{{A PCAC puzzle: $\pi^0 \to \gamma \gamma$ in the
  $\sigma$ model}}, \href{https://doi.org/10.1007/BF02823296}{\emph{Nuovo Cim.
  A} {\bfseries 60} (1969) 47}.

\bibitem{GAUME}
L.~Alvarez-Gaumé and E.~Witten, \emph{Gravitational anomalies},
  \href{https://doi.org/https://doi.org/10.1016/0550-3213(84)90066-X}{\emph{Nuclear
  Physics B} {\bfseries 234} (1984) 269 }.

\bibitem{AdlerBardeen}
S.~L. Adler and W.~A. Bardeen, \emph{Absence of higher-order corrections in the
  anomalous axial-vector divergence equation},
  \href{https://doi.org/10.1103/PhysRev.182.1517}{\emph{Phys. Rev.} {\bfseries
  182} (1969) 1517}.

\bibitem{Kharzeev_2016}
D.~Kharzeev, J.~Liao, S.~Voloshin and G.~Wang, \emph{Chiral magnetic and
  vortical effects in high-energy nuclear collisions—a status report},
  \href{https://doi.org/10.1016/j.ppnp.2016.01.001}{\emph{Progress in Particle
  and Nuclear Physics} {\bfseries 88} (2016) 1–28}.

\bibitem{Vazifeh}
M.~M. Vazifeh and M.~Franz, \emph{Electromagnetic response of weyl semimetals},
  \href{https://doi.org/10.1103/PhysRevLett.111.027201}{\emph{Phys. Rev. Lett.}
  {\bfseries 111} (2013) 027201}.

\bibitem{vilenkin}
A.~Vilenkin, \emph{Macroscopic parity-violating effects: Neutrino fluxes from
  rotating black holes and in rotating thermal radiation},
  \href{https://doi.org/10.1103/PhysRevD.20.1807}{\emph{Phys. Rev. D}
  {\bfseries 20} (1979) 1807}.

\bibitem{plasmas}
S.~Anand, J.~R. Bhatt and A.~K. Pandey, \emph{Chiral battery, scaling laws and
  magnetic fields},
  \href{https://doi.org/10.1088/1475-7516/2017/07/051}{\emph{Journal of
  Cosmology and Astroparticle Physics} {\bfseries 2017} (2017) 051–051}.

\bibitem{qcd}
D.~E. Kharzeev, \emph{The chiral magnetic effect and anomaly-induced
  transport}, \href{https://doi.org/10.1016/j.ppnp.2014.01.002}{\emph{Progress
  in Particle and Nuclear Physics} {\bfseries 75} (2014) 133–151}.

\bibitem{heavy}
D.~Kharzeev, J.~Liao, S.~Voloshin and G.~Wang, \emph{Chiral magnetic and
  vortical effects in high-energy nuclear collisions—a status report},
  \href{https://doi.org/10.1016/j.ppnp.2016.01.001}{\emph{Progress in Particle
  and Nuclear Physics} {\bfseries 88} (2016) 1–28}.

\bibitem{Burkov_2015}
A.~A. Burkov, \emph{Chiral anomaly and transport in weyl metals},
  \href{https://doi.org/10.1088/0953-8984/27/11/113201}{\emph{Journal of
  Physics: Condensed Matter} {\bfseries 27} (2015) 113201}.

\bibitem{Hosur_2013}
P.~Hosur and X.~Qi, \emph{Recent developments in transport phenomena in weyl
  semimetals}, \href{https://doi.org/10.1016/j.crhy.2013.10.010}{\emph{Comptes
  Rendus Physique} {\bfseries 14} (2013) 857–870}.

\bibitem{Kharzeev_2018}
D.~E. Kharzeev, Y.~Kikuchi and R.~Meyer, \emph{Chiral magnetic effect without
  chirality source in asymmetric weyl semimetals},
  \href{https://doi.org/10.1140/epjb/e2018-80418-1}{\emph{The European Physical
  Journal B} {\bfseries 91} (2018) }.

\bibitem{negative}
Q.~Li, D.~E. Kharzeev, C.~Zhang, Y.~Huang, I.~Pletikosić, A.~V. Fedorov
  et~al., \emph{Chiral magnetic effect in zrte5},
  \href{https://doi.org/10.1038/nphys3648}{\emph{Nature Physics} {\bfseries 12}
  (2016) 550–554}.

\bibitem{Gooth_2017}
J.~Gooth, A.~C. Niemann, T.~Meng, A.~G. Grushin, K.~Landsteiner, B.~Gotsmann
  et~al., \emph{Experimental signatures of the mixed axial–gravitational
  anomaly in the weyl semimetal nbp},
  \href{https://doi.org/10.1038/nature23005}{\emph{Nature} {\bfseries 547}
  (2017) 324–327}.

\bibitem{atoms}
X.-G. Huang, \emph{Simulating chiral magnetic and separation effects with
  spin-orbit coupled atomic gases},
  \href{https://doi.org/10.1038/srep20601}{\emph{Scientific Reports} {\bfseries
  6} (2016) }.

\bibitem{instab}
Y.~Akamatsu and N.~Yamamoto, \emph{Chiral plasma instabilities},
  \href{https://doi.org/10.1103/physrevlett.111.052002}{\emph{Physical Review
  Letters} {\bfseries 111} (2013) }.

\bibitem{tuchin2019timeevolution}
K.~Tuchin, \emph{Time-evolution of magnetic field in hot nuclear matter with
  fluctuating topological charge},  2019.

\bibitem{PhysRevD.92.074018}
C.~Manuel and J.~M. Torres-Rincon, \emph{Dynamical evolution of the chiral
  magnetic effect: Applications to the quark-gluon plasma},
  \href{https://doi.org/10.1103/PhysRevD.92.074018}{\emph{Phys. Rev. D}
  {\bfseries 92} (2015) 074018}.

\bibitem{PhysRevD.94.025009}
P.~V. Buividovich and M.~V. Ulybyshev, \emph{Numerical study of chiral plasma
  instability within the classical statistical field theory approach},
  \href{https://doi.org/10.1103/PhysRevD.94.025009}{\emph{Phys. Rev. D}
  {\bfseries 94} (2016) 025009}.

\bibitem{PhysRevD.92.125031}
Y.~Hirono, D.~E. Kharzeev and Y.~Yin, \emph{Self-similar inverse cascade of
  magnetic helicity driven by the chiral anomaly},
  \href{https://doi.org/10.1103/PhysRevD.92.125031}{\emph{Phys. Rev. D}
  {\bfseries 92} (2015) 125031}.

\bibitem{Avdoshkin_2016}
A.~{Avdoshkin}, V.~P. {Kirilin}, A.~V. {Sadofyev} and V.~I. {Zakharov},
  \emph{{On consistency of hydrodynamic approximation for chiral media}},
  \href{https://doi.org/10.1016/j.physletb.2016.01.048}{\emph{Physics Letters
  B} {\bfseries 755} (2016) 1}.

\bibitem{Miransky_2015}
V.~A. Miransky and I.~A. Shovkovy, \emph{Quantum field theory in a magnetic
  field: From quantum chromodynamics to graphene and dirac semimetals},
  \href{https://doi.org/10.1016/j.physrep.2015.02.003}{\emph{Physics Reports}
  {\bfseries 576} (2015) 1–209}.

\bibitem{wave}
D.~E. Kharzeev and H.-U. Yee, \emph{Chiral magnetic wave},
  \href{https://doi.org/10.1103/physrevd.83.085007}{\emph{Physical Review D}
  {\bfseries 83} (2011) }.

\bibitem{Shovkovy_2019}
I.~A. Shovkovy, D.~Rybalka and E.~Gorbar, \emph{The overdamped chiral magnetic
  wave}, \href{https://doi.org/10.22323/1.336.0029}{\emph{Proceedings of XIII
  Quark Confinement and the Hadron Spectrum — PoS(Confinement2018)} (2019) }.

\bibitem{Mottola:2019nui}
E.~Mottola and A.~V. Sadofyev, \emph{{Chiral Waves on the Fermi-Dirac Sea:
  Quantum Superfluidity and the Axial Anomaly}},
  \href{https://arxiv.org/abs/1909.01974}{{\ttfamily 1909.01974}}.

\bibitem{surowka}
D.~T. Son and P.~Surówka, \emph{Hydrodynamics with triangle anomalies},
  \href{https://doi.org/10.1103/physrevlett.103.191601}{\emph{Physical Review
  Letters} {\bfseries 103} (2009) }.

\bibitem{Sadofyev:2010pr}
A.~Sadofyev and M.~Isachenkov, \emph{{The Chiral magnetic effect in
  hydrodynamical approach}},
  \href{https://doi.org/10.1016/j.physletb.2011.02.041}{\emph{Phys. Lett. B}
  {\bfseries 697} (2011) 404}
  [\href{https://arxiv.org/abs/1010.1550}{{\ttfamily 1010.1550}}].

\bibitem{oz}
Y.~Neiman and Y.~Oz, \emph{Relativistic hydrodynamics with general anomalous
  charges}, \href{https://doi.org/10.1007/jhep03(2011)023}{\emph{Journal of
  High Energy Physics} {\bfseries 2011} (2011) }.

\bibitem{Landsteiner_2011}
K.~Landsteiner, E.~Megías, L.~Melgar and F.~Pena-Benitez, \emph{Holographic
  gravitational anomaly and chiral vortical effect},
  \href{https://doi.org/10.1007/jhep09(2011)121}{\emph{Journal of High Energy
  Physics} {\bfseries 2011} (2011) }.

\bibitem{Stone_2018}
M.~Stone and J.~Kim, \emph{Mixed anomalies: Chiral vortical effect and the
  sommerfeld expansion},
  \href{https://doi.org/10.1103/PhysRevD.98.025012}{\emph{Phys. Rev. D}
  {\bfseries 98} (2018) 025012}.

\bibitem{Jensen_2013}
K.~Jensen, R.~Loganayagam and A.~Yarom, \emph{Thermodynamics, gravitational
  anomalies and cones},
  \href{https://doi.org/10.1007/jhep02(2013)088}{\emph{Journal of High Energy
  Physics} {\bfseries 2013} (2013) }.

\bibitem{Golkar_2015}
S.~Golkar and D.~T. Son, \emph{(non)-renormalization of the chiral vortical
  effect coefficient},
  \href{https://doi.org/10.1007/jhep02(2015)169}{\emph{Journal of High Energy
  Physics} {\bfseries 2015} (2015) }.

\bibitem{Hou_2012}
D.-f. Hou, H.~Liu and H.-c. Ren, \emph{Possible higher order correction to the
  chiral vortical conductivity in a gauge field plasma},
  \href{https://doi.org/10.1103/physrevd.86.121703}{\emph{Physical Review D}
  {\bfseries 86} (2012) }.

\bibitem{Glorioso_2019}
P.~Glorioso, H.~Liu and S.~Rajagopal, \emph{Global anomalies, discrete
  symmetries and hydrodynamic effective actions},
  \href{https://doi.org/10.1007/jhep01(2019)043}{\emph{Journal of High Energy
  Physics} {\bfseries 2019} (2019) }.

\bibitem{PhysRevD.98.096011}
A.~Flachi and K.~Fukushima, \emph{Chiral vortical effect with finite rotation,
  temperature, and curvature},
  \href{https://doi.org/10.1103/PhysRevD.98.096011}{\emph{Phys. Rev. D}
  {\bfseries 98} (2018) 096011}.

\bibitem{Avkhadiev:2017fxj}
A.~Avkhadiev and A.~V. Sadofyev, \emph{{Chiral Vortical Effect for Bosons}},
  \href{https://doi.org/10.1103/PhysRevD.96.045015}{\emph{Phys. Rev. D}
  {\bfseries 96} (2017) 045015}
  [\href{https://arxiv.org/abs/1702.07340}{{\ttfamily 1702.07340}}].

\bibitem{Prokhorov:2020okl}
G.~Prokhorov, O.~Teryaev and V.~Zakharov, \emph{{CVE for photons: black-hole
  vs. flat-space derivation}},
  \href{https://arxiv.org/abs/2003.11119}{{\ttfamily 2003.11119}}.

\bibitem{Meg_as_2013}
E.~Megías and F.~Pena-Benitez, \emph{Holographic gravitational anomaly in
  first and second order hydrodynamics},
  \href{https://doi.org/10.1007/jhep05(2013)115}{\emph{Journal of High Energy
  Physics} {\bfseries 2013} (2013) }.

\bibitem{Bu:2015ika}
Y.~Bu and M.~Lublinsky, \emph{{Linearly resummed hydrodynamics in a weakly
  curved spacetime}},
  \href{https://doi.org/10.1007/JHEP04(2015)136}{\emph{JHEP} {\bfseries 04}
  (2015) 136} [\href{https://arxiv.org/abs/1502.08044}{{\ttfamily
  1502.08044}}].

\bibitem{Ma_es_2013}
J.~L. Mañes and M.~Valle, \emph{Parity violating gravitational response and
  anomalous constitutive relations},
  \href{https://doi.org/10.1007/jhep01(2013)008}{\emph{Journal of High Energy
  Physics} {\bfseries 2013} (2013) }.

\bibitem{Sadofyev_2018}
A.~Sadofyev and S.~Sen, \emph{Chiral anomalous dispersion},
  \href{https://doi.org/10.1007/jhep02(2018)099}{\emph{Journal of High Energy
  Physics} {\bfseries 2018} (2018) }.

\bibitem{Landsteiner:2013aba}
K.~Landsteiner, E.~Megias and F.~Pena-Benitez, \emph{{Frequency dependence of
  the Chiral Vortical Effect}},
  \href{https://doi.org/10.1103/PhysRevD.90.065026}{\emph{Phys. Rev. D}
  {\bfseries 90} (2014) 065026}
  [\href{https://arxiv.org/abs/1312.1204}{{\ttfamily 1312.1204}}].

\bibitem{Bu_2019}
Y.~Bu, T.~Demircik and M.~Lublinsky, \emph{Gradient resummation for nonlinear
  chiral transport: an insight from holography},
  \href{https://doi.org/10.1140/epjc/s10052-019-6576-z}{\emph{The European
  Physical Journal C} {\bfseries 79} (2019) }.

\bibitem{Maldacena_1999}
J.~Maldacena\href{https://doi.org/10.1023/a:1026654312961}{\emph{International
  Journal of Theoretical Physics} {\bfseries 38} (1999) 1113–1133}.

\bibitem{Yee_2010}
B.~Sahoo and H.-U. Yee, \emph{Holographic chiral shear waves from anomaly},
  \href{https://doi.org/10.1016/j.physletb.2010.04.076}{\emph{Physics Letters
  B} {\bfseries 689} (2010) 206–212}.

\bibitem{Erdmenger_2009}
J.~Erdmenger, M.~Haack, M.~Kaminski and A.~Yarom, \emph{Fluid dynamics of
  r-charged black holes},
  \href{https://doi.org/10.1088/1126-6708/2009/01/055}{\emph{Journal of High
  Energy Physics} {\bfseries 2009} (2009) 055–055}.

\bibitem{Kovtun_2012}
P.~Kovtun, \emph{Lectures on hydrodynamic fluctuations in relativistic
  theories},
  \href{https://doi.org/10.1088/1751-8113/45/47/473001}{\emph{Journal of
  Physics A: Mathematical and Theoretical} {\bfseries 45} (2012) 473001}.

\bibitem{Cvetic:1999ne}
M.~Cvetic and S.~S. Gubser, \emph{{Phases of R charged black holes, spinning
  branes and strongly coupled gauge theories}},
  \href{https://doi.org/10.1088/1126-6708/1999/04/024}{\emph{JHEP} {\bfseries
  04} (1999) 024} [\href{https://arxiv.org/abs/hep-th/9902195}{{\ttfamily
  hep-th/9902195}}].

\bibitem{Bhattacharyya:2008jc}
S.~Bhattacharyya, V.~E. Hubeny, S.~Minwalla and M.~Rangamani, \emph{{Nonlinear
  Fluid Dynamics from Gravity}},
  \href{https://doi.org/10.1088/1126-6708/2008/02/045}{\emph{JHEP} {\bfseries
  02} (2008) 045} [\href{https://arxiv.org/abs/0712.2456}{{\ttfamily
  0712.2456}}].

\end{thebibliography}\endgroup

\newpage

\appendix

\section{Boundary stress-energy tensor}\label{section:appB}

\indent  \indent Here we derive a convenient representation of Eq. (\ref{TEM}) in terms of the metric perturbations. This representation will be used in Section \ref{subsection:constitutive}  to determine the boundary expectation values of the curretns and stress-nergy tensor after find the solutions to the Einstein-Maxwell equations.

First, we define $r$-dependent stress energy tensor $\hat{T}(r,x)$ as
\begin{equation}\label{TEM1}
\hat{T}_{\mu \nu}(x,r)=  2 r^2\left( K_{\mu \nu}-K \gamma_{\mu \nu}-3 \gamma_{\mu \nu}- \frac{1}{2} \hat{G}_{\mu \nu}\right) 
+2 \bar\lambda \epsilon_{(\mu \alpha \beta \rho}F^{\alpha \beta} R_{\nu)}^{\rho}.
\end{equation}

With that (\ref{TEM1}) looks like
\begin{equation}
T_{\mu \nu}(x)=\frac{1}{16 \pi G_5} \lim\limits_{r \rightarrow \infty}\left[ \hat{T}_{\mu \nu}(x,r)+T^{ct}_{\mu \nu}(x,r) \log \left( \frac{1}{r^2}\right)\right]
\end{equation}

Expressing Eq. (\ref{TEM}) in terms of metric  Eq. (\ref{metrdec}) and linearizing in the perturbation amplitude, one obtains the following explicit expression for the $r$-depenedent stress-energy tensor \cite{Bu:2015ika}:

\begin{subequations}\label{SET}
\begin{gather}\label{stress tensor1}
\hat{T}_{00}=3-3 h_{00}^{CFT}
+ \left[ \frac{1}{2}r^2 \partial_i\partial_j h_{ij}^{CFT}-\frac{1}{2}r^2\partial^2 h_{kk}^{CFT}-r^2\partial^2h+\frac{1}{2}r^2\partial_i\partial_j \pi_{ij} -9r^4h+\right. \nonumber \\
 \left. +3k-2r^3\partial u+2r^3\partial_k h_{0k}^{CFT}-r^3\partial_t h_{kk}^{CFT}+ 2 r^3 \partial j-3r^3\partial_t h-3r^5\partial_r h \right].
\end{gather}
\begin{gather}
\hat{T}_{0i}=-4 u_i+ h_{0i}^{CFT}
+\left[\frac{1}{2}r^3\partial_i h_{00}^{CFT}+\frac{1}{r}\partial_ik-r^5 \partial_r j_i-r^3\partial_t u_i-\frac{3}{2}r^3\partial_ih  \right.\nonumber\\
\left.-\frac{1}{2}r^2\partial^2h_{0i}^{CFT}+\frac{1}{2}r^2\partial_i\partial_k h_{0k}^{CFT} + \frac{1}{2}r^2\partial_t\partial_k h_{ik}^{CFT}-\frac{1}{2}r^2\partial_t\partial_i h_{kk}^{CFT}
- \frac{1}{2}r^2\partial^2 j_i+\right. \nonumber\\
\left.+\frac{1}{2}r^2\partial_i\partial j+\frac{1}{2}r^2\partial_t\partial_k \pi_{ik} -r^2\partial_t\partial_ih\right]
\end{gather}
\begin{gather}
\hat{T}_{ij}=\delta_{ij}+ h_{ij}^{CFT}+ \delta_{ij} \left[-\frac{1}{2}r^2\partial^2h_{00}^{CFT}+ \frac{1}{2}r^2\partial^2h_{kk}^{CFT}- \frac{1}{2} r^2\partial_k\partial_lh_{kl}^{CFT}+r^2\partial_t\partial_k h_{0k}^{CFT}-\right.\nonumber\\
\left. -\frac{1}{2}r^2 \partial_t^2 h_{kk}^{CFT}+ \frac{1}{2}r^2\partial^2h-\frac{1}{2r^2} \partial^2k-\frac{r^2}{2} \partial_k\partial_l\pi_{kl}+r^2\partial_t\partial j-\right. \nonumber\\
\left.-r^2\partial_t^2h+9r^4h+k+2r^3\partial u-2r^3\partial_kh_{0k}^{CFT}+ r^3\partial_t h_{kk}^{CFT}-2 r^3\partial j -r^3\partial_t h\right. \nonumber\\
\left.+ \frac{1}{r}\partial_t k +2r^5\partial_r h-r\partial_r k\right]
+\left[\frac{1}{2}r^2\partial_i\partial_jh_{00}^{CFT}-\frac{1}{2}r^2\partial^2h_{ij}^{CFT} -\frac{1}{2}r^2\partial_i\partial_j h_{kk}^{CFT}+\right. \nonumber\\
\left.+\frac{1}{2}r^2\left(\partial_i\partial_kh_{jk}^{CFT}+ \partial_j\partial_kh_{ik}^{CFT}\right)-\frac{1}{2}r^2\partial_t\left(\partial_ih_{0j}^{CFT}+ \partial_jh_{0i}^{CFT} \right) +\frac{1}{2}r^2\partial_v^2 h_{ij}^{CFT}-\right. \nonumber \\
\left.-\frac{1}{2}r^2 \partial_i\partial_jh+ \frac{1}{2r^2}\partial_i\partial_jk- \frac{1}{2}r^2 \partial^2\pi_{ij}
+\frac{1}{2} r^2\left(\partial_i\partial_k\pi_{jk}+\partial_j\partial_k \pi_{ik}\right) \right. \nonumber\\
\left.-\frac{1}{2}r^2 \partial_t\left(\partial_ij_j+\partial_j j_i\right) +\frac{1}{2}r^2 \partial_t^2\pi_{ij}-r^3\left(\partial_iu_j+ \partial_j u_i\right)+r^3\left(\partial_i h_{0j}^{CFT} + \partial_j h_{0i}^{CFT}\right)  \right. \nonumber\\
\left.-r^3\partial_t h_{ij}^{CFT}+ r^3 \left(\partial_i j_j +\partial_j j_i\right)-r^3 \partial_t \pi_{ij} - r^5\partial_r\pi_{ij}\right] ,
\end{gather}
\end{subequations}
where we neglected the terms vanishing at the boundary.

Eqs. (\ref{SET}) can be considerably simplified if we restrict ourselves to the only relevant for us  helicity $\pm1$ and $\pm 2$ sectors.  

Going to the Fourier space, for helicity $\pm 1$ perturbations we have:

\begin{subequations}\label{app1}
\begin{gather}
 \hat{T}_{00}(k)=3
\end{gather}\begin{gather}
\hat{T}_{01}(k)=-4u_1+h_{01}^{CFT}- r^5 \partial_r j_1
+\frac{1}{2}r^2 k^2 h_{01}^{CFT}+\frac{1}{2}r^2 k^2 j_1
\end{gather}\begin{equation}
\hat{T}_{13}(k)=h_{13}^{CFT}-r^3 i k u_1+r^3i k h_{01}^{CFT}+r^3 i k j_1 -r^5 \partial_r \pi_{13}
\end{equation}

\end{subequations}

Expressions for $T_{23}(k)$ and $T_{02}(k)$ are the same as those for $T_{13}$ and $T_{02}$, correspondingly, provided we exchange the indices $1 \leftrightarrow 2$.

Similarly, for helicity $\pm2$ perturbations the stress-energy tensor is
\begin{subequations}\label{app2}
\begin{equation}
\hat{T}_{00}(k)=3,
\end{equation}
\begin{equation}
\bar{T}_{0i}(k)=-4 u_i,
\end{equation}
\begin{equation}
\hat{T}_{ij}(k)=\delta_{ij}+h_{ij}+\frac{1}{2}r^2 k^2 h_{ij}+\frac{1}{2}r^2 k^2 \pi_{ij}-r^5 \partial_r \pi_{ij}.
\end{equation}
\end{subequations}
Eqns. (\ref{app1},\ref{app2}) are valid to all orders in $k$, but only to the first order in perturbations.
\end{document}